\documentclass[manuscript]{aastex}

\begin{document}


\title{SDSS J124602.54+011318.8: A Highly Variable AGN, Not
an Orphan GRB Afterglow}


\author{Avishay Gal-Yam\altaffilmark{1}, Eran O. Ofek}
\affil{School of Physics \& Astronomy, Tel-Aviv University, Tel-Aviv
69978, Israel}
\email{avishay@wise.tau.ac.il}

\and

\author{Alexei V. Filippenko, Ryan Chornock, and
Weidong Li}
\affil{Department of Astronomy, 601 Campbell Hall, University of
California, Berkeley, CA 94720-3411, USA}


\altaffiltext{1}{Colton Fellow.}


\begin{abstract}

The optically variable source SDSS J124602.54+011318.8 first appears in Sloan
Digital Sky Survey (SDSS) data as a bright point source with nonstellar
colors. Subsequent SDSS imaging and spectroscopy showed that the point source
declined or disappeared, revealing an underlying host galaxy at redshift
0.385. Based on these properties, the source was suggested to be a candidate
``orphan afterglow'': a moderately beamed optical transient, associated with a
gamma-ray burst (GRB) whose highly beamed radiation cone does not include our
line of sight. We present new imaging and spectroscopic observations of this
source.  When combined with a careful re-analysis of archival optical and radio
data, the observations prove that SDSS J124602.54+011318.8 is in fact an
unusual radio-loud AGN, probably in the BL Lac class. The object displays
strong photometric variability on time scales of weeks to years, including
several bright flares, similar to the one initially reported. The SDSS
observations are therefore almost certainly not related to a GRB.  The optical
spectrum of this object dramatically changes in correlation with its optical
brightness.  At the bright phase, weak, narrow oxygen emission lines and
probably a broader H$\alpha$ line are superposed on a blue continuum.  As the
flux decreases, the spectrum becomes dominated by the host galaxy light, with
emerging stellar absorption lines, while both the narrow and broad emission
lines have larger equivalent widths.  We briefly discuss the implications of
this discovery on the study of AGNs and other optically variable or transient
phenomena.

\end{abstract}




\section{Discovery and Basic Properties}

SDSS J124602.54+011318.8 was discovered in Sloan Digital Sky Survey (SDSS) data
obtained during March 1999. Full details are given by Vanden Berk et al.
(2001; hereafter VB01). Being an $r' \approx 17$ mag point source, with
nonstellar colors and a radio counterpart from the FIRST survey, the object was
flagged as a candidate active galactic nucleus (AGN). Subsequent SDSS imaging
and spectroscopy during April--May 2000 and January 2001 revealed that the
source had faded considerably to $r' \approx 19.5$ mag, became marginally
resolved, and appeared to have a normal galactic spectrum.

These remarkable changes triggered the subsequent study by VB01.  From the SDSS
data available to VB01, the source appeared as a bright optical transient (OT)
that erupted during 1999 and later declined, revealing the underlying host
galaxy in spectra and images obtained over a year later. VB01 suggest that the
object is indeed an OT similar to ones associated with gamma-ray bursts (GRBs),
but do not rule out the possibility that this object is an unusual AGN.  Within
the astrometric errors of the SDSS images, the location of the point source is
consistent with the galaxy nucleus.  Combining the SDSS data with archival
images from the Digitized Sky Survey, VB01 claim that the source shows no
evidence for past variability, and is therefore less likely to be a highly
variable AGN.  They advocate future optical monitoring to further check this
issue.

Following the suggestion of VB01, we observed SDSS J124602.54+011318.8 in the
$R$ band with the Wise Observatory 1-m telescope. Combining our new data, the
results reported by VB01, and a careful re-analysis of publicly available
archival images of the field, we find that SDSS J124602.54+011318.8 is in fact
a highly variable object, and not a transient phenomenon.  As our January 2002
photometry showed that this object was in a bright ($R \approx 18$ mag) state,
we have obtained optical spectra of SDSS J124602.54+011318.8 using the Shane
3-m telescope at Lick Observatory and the Keck-I 10-m telescope in Hawaii. In
striking contrast to the data presented by VB01, the spectra show a blue
continuum with weak broad lines, suggestive of an AGN.  The redshift ($z =
0.385$), measured from narrow nebular emission lines, is identical to the one
found by VB01 from the galactic spectrum they obtained. The combination of the
strong optical variability and the spectral information leads us to conclude
that SDSS J124602.54+011318.8 is indeed an unusual radio-loud AGN that might be
classified as a blazar, probably akin to the BL Lac subclass. Since this object
displayed several flares similar to the one observed by the SDSS, there is no
reason to suspect that a GRB, or any other sort of non-AGN optical transient,
has occurred in this active galaxy.

In the following sections, we present the analysis of the optical photometry
($\S~2$) and spectroscopy ($\S~3$) of this object, its radio and X-ray
properties ($\S~4$), and their implications ($\S~5$).
 



\section{Photometry}

Figure 1 shows an $R$-band image of the field of SDSS J124602.54+011318.8,
obtained on 10 January 2002 (UT dates are used throughout this paper) with the
Wise Observatory 1-m telescope.  The object, at $R \approx 18$ mag, is
marginally resolved. In Figure 2 we plot $R$-band magnitudes of this object
measured from new and archival data from 1956 to date. We have transformed SDSS
$r'$ magnitudes to Cousins $R$ using the prescription of Smith et al. (2002)
and have taken $E$ (red) magnitudes from Palomar Observatory Sky Survey (POSS)
plates to be equal to Cousins $R$, albeit with a large uncertainty of $\pm 0.2$
mag, following the analysis of Ofek (2000). Unfiltered observations from the
LONEOS photometric database (Rest et al. 2001) are tied to the $R$ band 
(A. Miceli, private communication). One can see that the object is
variable with a peak-to-peak amplitude of more than 2.5 mag. High states were
detected during 1991, 1999, and again in our new data. The SDSS images from
2000 and 2001 were apparently taken during a period of very low
activity. Examining the blue data from the SDSS, presented by VB01, and
analogous archival $O$ (blue) plates, we find a higher amplitude of variability
($\sim4$ mag), probably because the blue bands are dominated by the variable
AGN while the light from the host galaxy contributes more to the $R$ band.

We have analyzed archival unfiltered observations of this field obtained by the
NEAT project (Pravdo et al. 1999), and made available through the {\it
skymorph}\footnotemark[2] website. Data of good quality were obtained during
February--April of 2000 and 2001.  Figure 3 (top panel) shows the unfiltered
magnitude of SDSS J124602.54+011318.8 relative to the magnitude of a nearby
bright star (marked as Star 1 in Fig. 1). One can see that the object shows
large-amplitude variations (peak to peak variability of more than 1 mag) over
time scales of weeks and months. Some variability on a time scale of days is
also evident, but needs to be verified by higher-quality photometry. We have
tested the reality of the measured variability by performing the same
measurement on a nearby star (marked as Star 2 in Fig. 1) with similar
flux. The results (plotted in the bottom panel of Fig. 3) show that the
measured flux is generally constant, with a dispersion of $\sim 0.05$ mag
around the mean, making the variability of SDSS J124602.54+011318.8 a
$20\sigma$ effect. Since the host galaxy of SDSS J124602.54+011318.8 may become
dominant when the AGN is weak, thus introducing seeing dependence into our flux
measurement, we repeated our photometry test with a nearby resolved galaxy
(marked as Galaxy 1 in Fig. 1). The results, also given at the bottom panel of
Figure 3, show that while the scatter in flux is indeed somewhat larger for
this object ($\sim 0.1$ mag), the variability of SDSS J124602.54+011318.8 is
still highly significant ($10\sigma$). We conclude that SDSS
J124602.54+011318.8 is beyond doubt a variable object, with large fluctuations
in flux on time scales of weeks to years.

\footnotetext[2]{\url{http://skys.gsfc.nasa.gov/skymorph/skymorph.html}} 

\section{Spectroscopy}

   We obtained optical spectra of SDSS J124602.54+011318.8 twice to assess its
spectral variability.  The first observation was a 45-minute exposure taken on
14.5 January 2002 with the Kast double spectrograph (Miller \& Stone 1993) on
the Shane 3-m telescope at Lick Observatory.  The second spectrum was a
5-minute observation taken on 17.7 January 2002 with the Low Resolution Imaging
Spectrometer (LRIS; Oke et al. 1995) mounted on the Keck-I 10-m telescope.  The
long, $2''$ wide Lick Kast slit and $1''$ wide Keck LRIS slit were both aligned
with the parallactic angle to minimize light losses due to atmospheric
dispersion (Filippenko 1982). The standard IRAF\footnotemark[3] routines were
\footnotetext[3]{IRAF is distributed by the National Optical Astronomy
Observatories, which are operated by the Association of Universities for
Research in Astronomy, Inc., under cooperative agreement with the National
Science Foundation.}
used for two-dimensional image processing and optimal spectral extraction
(Horne 1986).  We used our own IDL routines (Matheson et al. 2000) to do the
flux calibration and atmospheric band removal (Wade \& Horne 1988).  In each
case, the spectral range was 3500--10000~\AA. The Lick spectrum had an anomaly,
probably resulting from an imperfect join between its blue and red sides (whose
light is split with a dichroic filter), that prevented accurate continuum
calibration in the range 4800--6000~\AA, so that region has been excised from
Figure 4.

   Both spectra were taken in clear skies (some very thin cirrus may have been
present). We have derived $R$-band magnitudes by convolving the flux-calibrated
spectra with the $R$-band response function, but the associated uncertainties
could be only roughly estimated by considering the stability of the nights.
The results are $R = 16.9 \pm 0.3$ mag for the 14.5 January Lick spectrum (the
uncertainty is relatively large due to variable seeing), and $R = 18.1 \pm
0.15$ for the 17.7 January 2002 Keck spectrum. The Lick magnitude supports our
suspicion, based on spectral characteristics (see below), that the object was
substantially brighter than during the 12 January Wise observation ($R = 18.0$
mag) shown in the inset of Figure 2.  The Keck magnitude is very close to the
18 January Wise measurement obtained just 12 hours later.

The SDSS spectroscopy presented by VB01\footnotemark[4] was obtained while the
object was quiescent, and is therefore dominated by the light from the host
galaxy, resulting in a generally red spectrum with stellar absorption
lines. However, the narrow [O~III] emission lines and possible broad H$\alpha$
may suggest some AGN contribution, in accord with the radio
detection. Examining more than 200 spectra of radio-loud AGNs presented by
Stickel \& Kuehr (1994), Stickel et al. (1996), Kock et al. (1996), Nass et
al. (1996), Perlman et al. (1998), and Landt et al. (2001), we find only two
spectra (WGAJ0245.2+1047, Perlman et al. 1998; WGAJ0528.5-5820, Landt et
al. 2001) that resemble that of SDSS J124602.54+011318.8.  Both objects are
classified as BL Lacs, but the SDSS spectrum seems to be even more ``galactic''
(e.g., with stronger Ca~II H \& K lines), so it may in fact resemble that of a
radio galaxy.

\footnotetext[4]{Data are publicly available through the SDSS spectral archive,
\url{http://archive.stsci.edu/sdss/spectra.html .}}

In contrast, the nearly featureless Lick spectrum, shown in Figure 4, was
obtained when the object was in a bright phase (Fig. 2) and is apparently
dominated by a nonstellar source.  The continuum is very blue, with low
equivalent width emission lines of [O~III] and [O~II], and possibly broader
H$\alpha$ and Mg~II lines. The spectrum is very suggestive of a BL Lac
object. No stellar absorption lines from the host are visible, but the measured
redshift is $z=0.385$ from the narrow [O~III] line, matching the SDSS redshift.

The Keck LRIS spectrum, obtained $\sim 3$ days later and shown in Figure 5, was
apparently taken when the object was at an intermediate phase (being about one
third as bright as it was when the spectrum shown in Fig. 4 was taken).  The
continuum is less blue, and stellar absorption lines are visible, although they
are not as strong as in the SDSS spectra. The broad H$\alpha$ line is now
clearly visible, with equivalent width $(W_{\lambda}) \approx 13$~\AA\ and full
width at half maximum (FWHM) $\approx 3000$ km s$^{-1}$.  The emergence of
broad lines during the continuum drop that accompanies the transition to a low
activity state is well documented in variable radio-loud AGNs (e.g., Vermeulen
et al. 1995; Koratkar et al. 1998).  The strength of the ``4000~\AA\ break" is
$C \approx 0.1$, following the definition of Perlman et al. (1998).  Combining
our measured values for $C$ and $W_{\lambda}$ and using the classification
diagram of March$\tilde{{\rm a}}$ et al. (1996) and Perlman et al. (1998), we
find that this object falls on the outskirts of the BL Lac region.

Wise photometry taken $\sim 12$ hours after the Keck LRIS spectrum indeed shows
the object to be fainter than it was before the Lick spectrum was obtained, but
still brighter than it was at the time of the SDSS spectroscopy (Fig. 2). This
is consistent with the spectral analysis showing the Keck spectrum to be
intermediate between the AGN-dominated Lick spectrum and the host-dominated
SDSS spectra.

 In conclusion, SDSS J124602.54+011318.8 shows dramatic spectral variability,
apparently correlated with its optical flux.  The broad lines and the blue,
variable, nonstellar continuum observed in both the Lick and Keck spectra are
indicative of an AGN. The combination of the spectral data from all epochs is
best explained by an extreme, strongly variable BL-Lac-type object that
resembles a radio galaxy during its faint phase. Since the properties of the
flare reported by VB01 are fully replicated by our new observations of this
AGN, there is no reason to suspect that an OT occurred during March 1999.

\section{Radio and X-ray Properties}

The FIRST 1.4 GHz radio catalog (White et al. 1997) lists only one source
within $3'$ of SDSS J124602.54+011318.8, coincident with the SDSS position of
the optical source to $\pm 0.3''$, well below the combined astrometric errors
of the SDSS and FIRST data. Therefore, one can safely identify the $79.4 \pm
0.2$ mJy FIRST source with SDSS J124602.54+011318.8. Comparing the FIRST
detection with another 1.4 GHz survey, we find that the NVSS catalog (Condon et
al. 1998) quotes a significantly lower flux measurement of $71.9 \pm 2$ mJy.
Scanning other radio catalogs, we have found that this source was detected at
4.85 GHz by the PMN survey (Griffith \& Wright 1993; $51 \pm 11$ mJy) sometime
during 1990, while the GB6 catalog (Gregory et al. 1996; $90 \pm 11$ mJy)
detects the source at the same frequency, in data taken during 1986--1987
(exact epochs are not currently available). We therefore conclude that this
source is probably variable at radio wavelengths, with a change of up to a
factor of $\sim 2$ in flux over several years. Such radio properties are
consistent with the characteristics of radio-loud AGNs in general, and BL Lacs
in particular. The object is not detected in the ROSAT all-sky survey, nor does
it appear in the EGRET third catalog.

\section{Discussion}

The identification of SDSS J124602.54+011318.8 as an unusual AGN has several
implications in various related fields, including the study of optical
transients in general and optical afterglows of GRBs in particular, as well as
AGN research. We briefly comment on these subjects.

Perhaps the most important lesson to be learned from the work presented above
is the need for a thorough search of available archival data, when studying
potentially transient or variable objects. Large amounts of high-quality
astronomical data are now easily available to the community, covering most of
the sky in a wide range of wavelengths. In many cases, archival data can
provide evidence for past variability, or at least constrain it. There are many
peculiar stars and AGNs out there, and when large areas of the sky are
searched, they are bound to appear occasionally. Without proper care, such
objects may pose as transient events such as supernovae or GRB afterglows. Even
when archival data are not available or show no evidence for variability,
continued optical monitoring of the sites of candidate transients may be
prudent, if strong conclusions are to be drawn from such events. We note that
several theoretical papers on SDSS J124602.54+011318.8 have already been
submitted for publication (e.g., Granot et al.  2002), and although they may
still serve a useful purpose for general discussions, the specific conclusions
regarding this particular ``orphan GRB" clearly are not applicable.

One of the interesting applications of the accurate SDSS 5-band photometry is
the possible distinction between different classes of objects by their colors
(Krisciunas, Margon, \& Szkody 1998; Poznanski et al. 2002).  VB01 demonstrate
that the optical colors of SDSS J124602.54+011318.8 in the bright phase can be
fitted by a power law with index $\beta_{\nu} \approx 1$, and claim that these
colors are similar to those of GRB afterglows and inconsistent with those of
most AGNs. Our work shows that at least some AGNs have colors that are similar
to those of GRBs. In fact, such colors have been measured for well-known
flaring AGNs such as BL Lac (Webb et al. 1998) and the radio-loud quasar 3C~279
(Webb et al. 1990). The selection of GRB afterglows by their colors alone would
therefore seem difficult.  VB01 claim that the discovery of an ``orphan
afterglow'' within the amount of SDSS data searched so far disfavors the models
used in the calculations of Dalal, Griest, \& Pruet (2002), which predict very
few ``orphan afterglows'' even in deep, wide-field searches. The fact that SDSS
J124602.54+011318.8 probably did not host a GRB implies that these models are
as yet uncontested.

The strong spectral variability this object displays also has some implications
for studies of radio-loud AGNs. Looking only at low-state optical spectra of
SDSS J124602.54+011318.8, it is difficult to classify it as a blazar, since the
nonstellar continuum is well below the level of the host-galaxy light. The fact
that surveys for AGNs may miss a population of such low-luminosity blazars, for
exactly this reason, has already been noted (Browne \& March$\tilde{{\rm a}}$
1993; see a thorough recent discussion in Perlman et al. 1998). However,
besides being an extreme example of the ``Browne \& March$\tilde{{\rm a}}$
effect," SDSS J124602.54+011318.8 also raises the issue of variability. The
calculations presented by Browne \& March$\tilde{{\rm a}}$ (1993),
March$\tilde{{\rm a}}$ \& Browne (1995), and Perlman et al. (1998) do not take
the intrinsic variability of the sources into account. Objects similar to SDSS
J124602.54+011318.8 may spend a significant fraction of their time in a low
state, where they are hard to recognize, suffering an increased Browne \&
March$\tilde{{\rm a}}$ effect. In order to estimate how significant this might
be (i.e., the incompleteness of blazar surveys), one needs to determine the
duty cycle of objects like SDSS J124602.54+011318.8, as well as know their
relative frequency within the entire blazar population. Both of these
properties may be measured by a spectroscopic monitoring campaign of a complete
sample of radio-selected AGNs.

A similar aspect of the extreme variability of this object is the pronounced
change in its apparent optical morphology, from a bright, star-like point
source to a faint resolved galaxy. If such objects are not very rare, programs
using optical morphology to select candidate AGNs may be strongly affected. As
an example, Wallace et al.  (2002) searched for the blazar counterpart of the
unidentified EGRET source 3EG J2006-2321, using radio and optical data.  In
order to decide between the two most probable radio sources that fall within
the EGRET error contour, they examine deep optical images of this field. One of
the sources appears like ``a normal-looking galaxy'' while the other is
stellar. The authors use this to rule out the galactic-looking source. In this
particular case, the decision was probably well justified, as the remaining
source shows spectral and polarization properties that are characteristic of
blazars. But a population of objects similar to SDSS J124602.54+011318.8, with
variable apparent morphology, may hinder the use of object morphology as a
selection method in similar studies.

Finally, we note that SDSS J124602.54+011318.8 presents an opportunity to study
the properties of the stellar population in the nucleus of an AGN host
galaxy; when in the low state, the AGN is faint and does not obstruct the
observations of the host, unlike the case for its brighter counterparts.

\section*{Acknowledgments}

A. Miceli is thanked for providing us with the LONEOS data points.  We are
grateful for valuable advice from N. Brosch, S. Kaspi, A. Laor, T. Piran,
O. Shemmer, A. Sternberg, and C. M. Urry. Helpful comments on the manuscript
were provided by A. Levinson and H. Netzer. It is a pleasure to acknowledge the
referee, E. Perlman, for his helpful suggestions.  We acknowledge the
assistance of the staffs of the Wise, Lick, and Keck Observatories; Y.-J. Choi
and Y. Lipkin are especially thanked for their help at Wise. The W. M. Keck
Observatory is operated as a scientific partnership among the California
Institute of Technology, the University of California, and NASA; it was made
possible by the generous financial support of the W. M. Keck Foundation.
Astronomy at the Wise Observatory is supported by grants from the Israel
Science Foundation. A.V.F. is grateful for the financial support of NSF grant
AST-9987438, as well as of the Guggenheim Foundation.

\clearpage


\begin{figure*}
\plotone{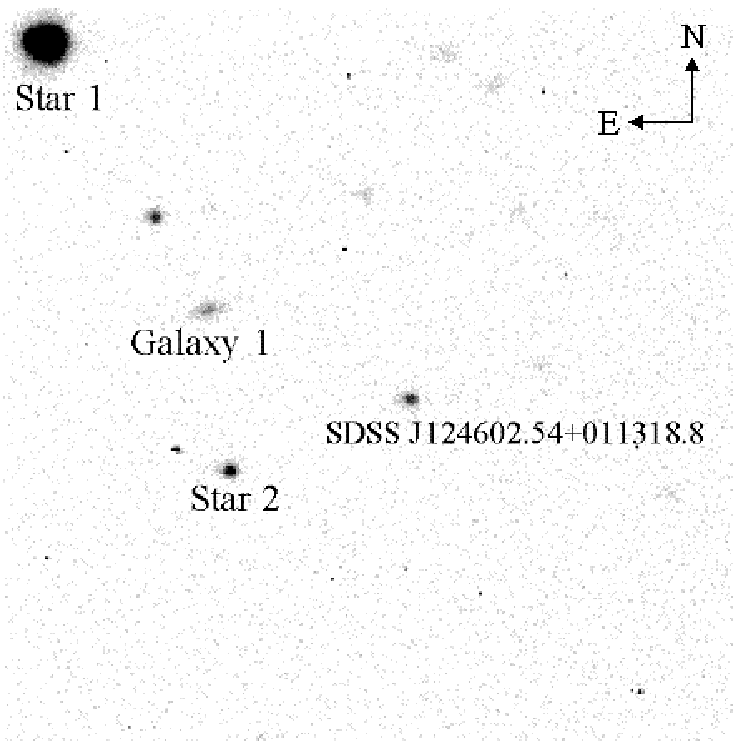}
\caption{A $3' \times 3'$ section of a 900-s $R$-band image obtained with the
Tek CCD camera mounted on the Wise Observatory 1-m telescope. SDSS
J124602.54+011318.8 and comparison objects are marked. Note that SDSS
J124602.54+011318.8 is marginally resolved.}
\end{figure*}

\begin{figure*}
\plotone{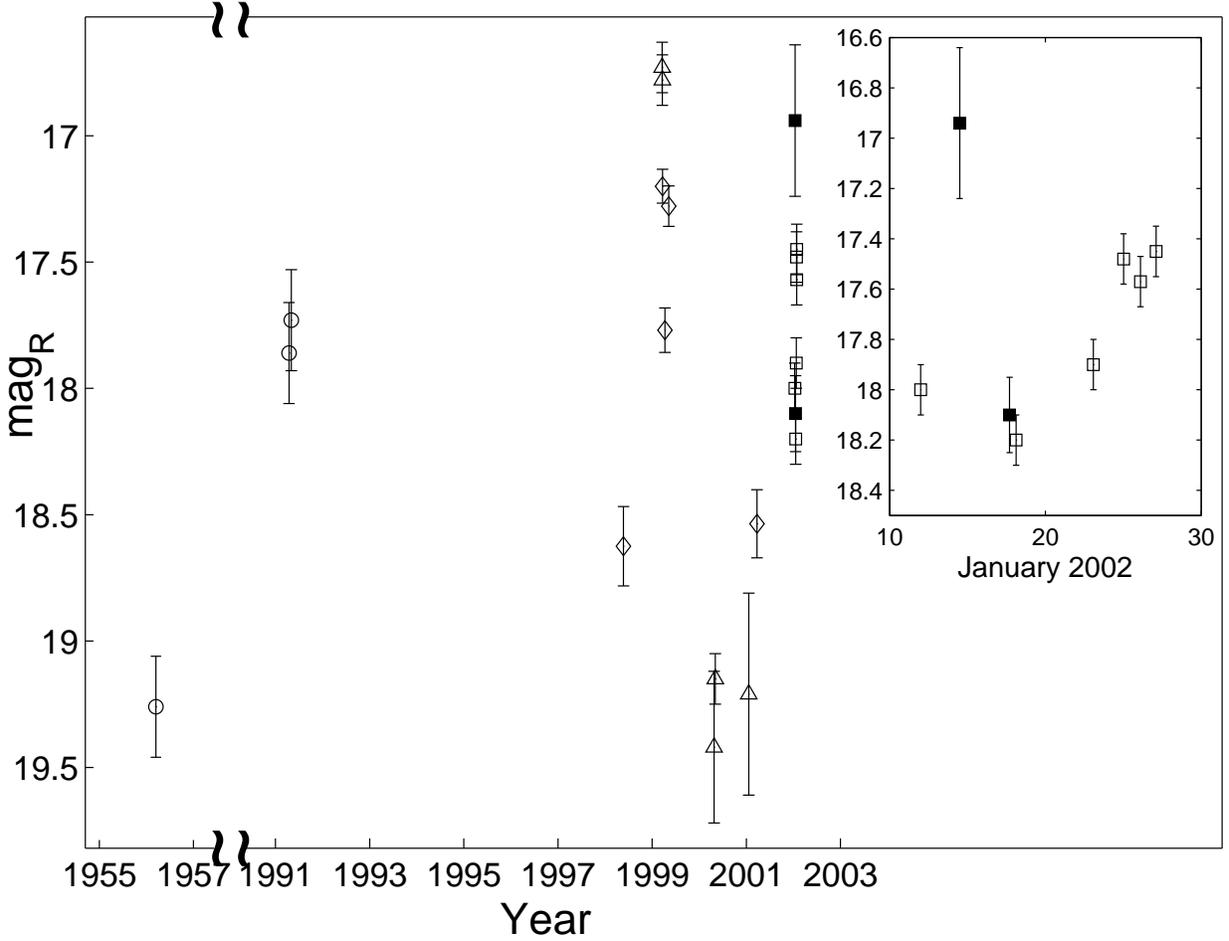}
\caption{$R$-band light curve of SDSS J124602.54+011318.8 spanning 45 years of
data. The combination of archival digital POSS plates ({\it circles}),
multi-epoch SDSS data from VB01 ({\it triangles}), archival LONEOS data ({\it
diamonds}), our new Wise data ({\it open squares}), and spectral 
flux measurements ({\it filled squares}; see \S~3) 
shows unambiguously that the source is strongly
variable. (Note that the error bars on the spectral measurements are not formal
$1\sigma$ values, but rather are estimates based on the perceived stability of
the night.) The object appears to have been quiescent ($R \approx 19.3$) on 
the 1956 POSS-I plate and during the SDSS spectroscopy and imaging runs in
2000--2001. During March-May 1999, SDSS and LONEOS observations apparently
caught the object in a high state, while the 1991 POSS-II data and our Wise
observations are intermediate, but still significantly higher than the
low-state baseline. The inset shows a detailed view of the recent Wise
photometry, along with the spectral flux measurements.}
\end{figure*}

\begin{figure*}
\plotone{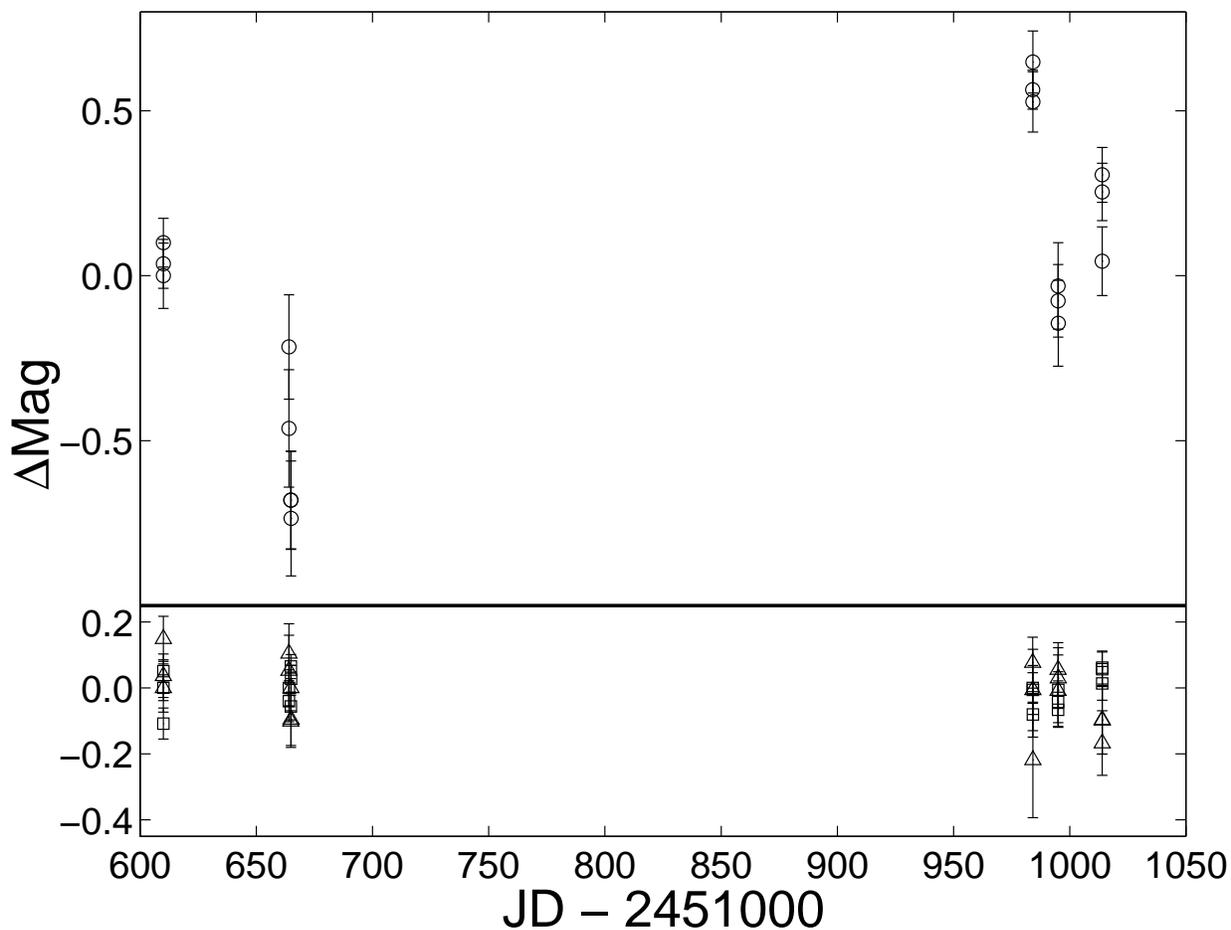}
\caption{{\bf Upper panel:} The unfiltered magnitudes of SDSS
J124602.54+011318.8 from NEAT images, as a function of time, relative to the
median. Each data point was normalized by the measured flux of a nearby bright
star (marked as Star 1 in Fig. 1), assumed to be stable.  {\bf Lower panel:}
The same as the top panel, but for a star with flux similar to that of SDSS
J124602.54+011318.8 (Star 2 in Fig. 1; {\it open squares}) and a nearby galaxy
(Galaxy 1 in Fig. 1; {\it open triangles}). For both objects the measurements
are constant within the errors. The data show that SDSS J124602.54+011318.8
varies on time scales of years, months, and weeks.}
\end{figure*}

\begin{figure*}
\plotone{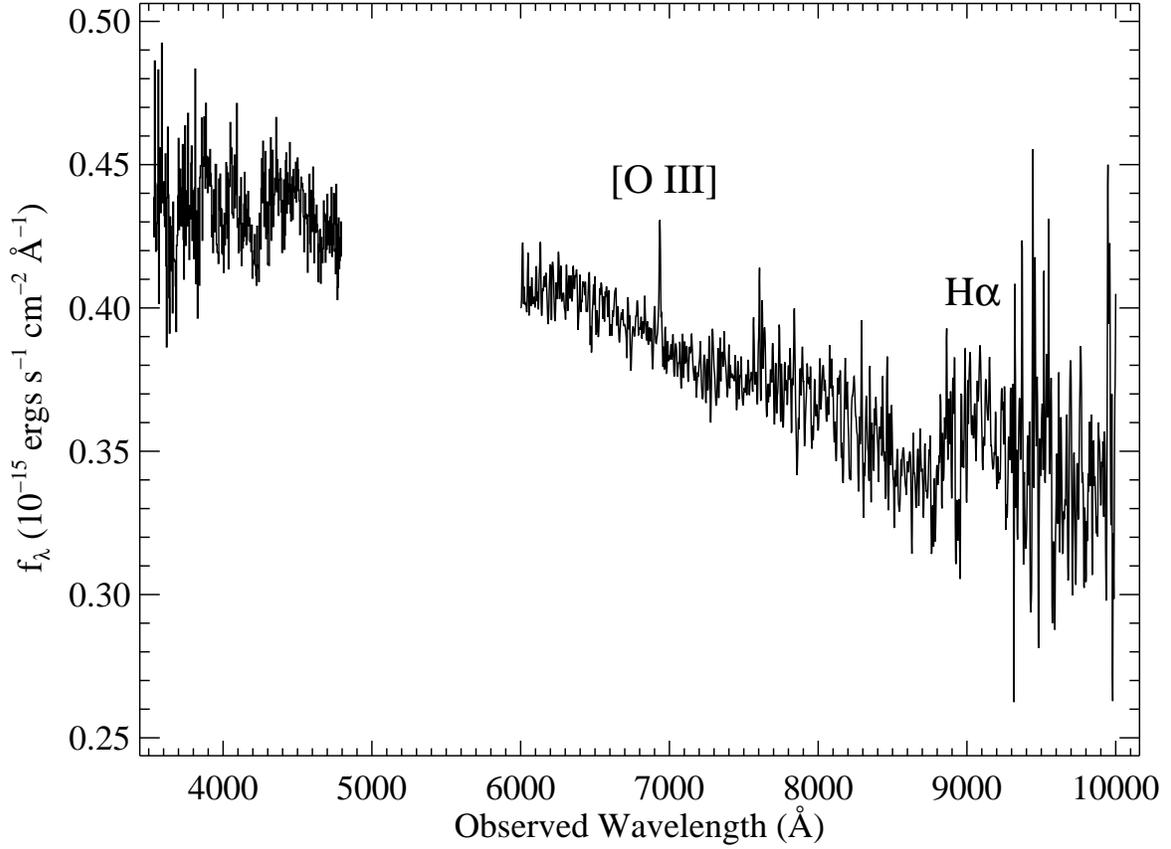}
\caption{A spectrum of SDSS J124602.54+011318.8 obtained on 14.5 January 2002
with the 3-m Shane reflector at Lick Observatory. Note the blue, featureless
continuum and the probable weak, broad H$\alpha$ emission line. The redshift
measured from the narrow [O~III] $\lambda$5007 line is 0.385.}
\end{figure*}

\begin{figure*}
\plotone{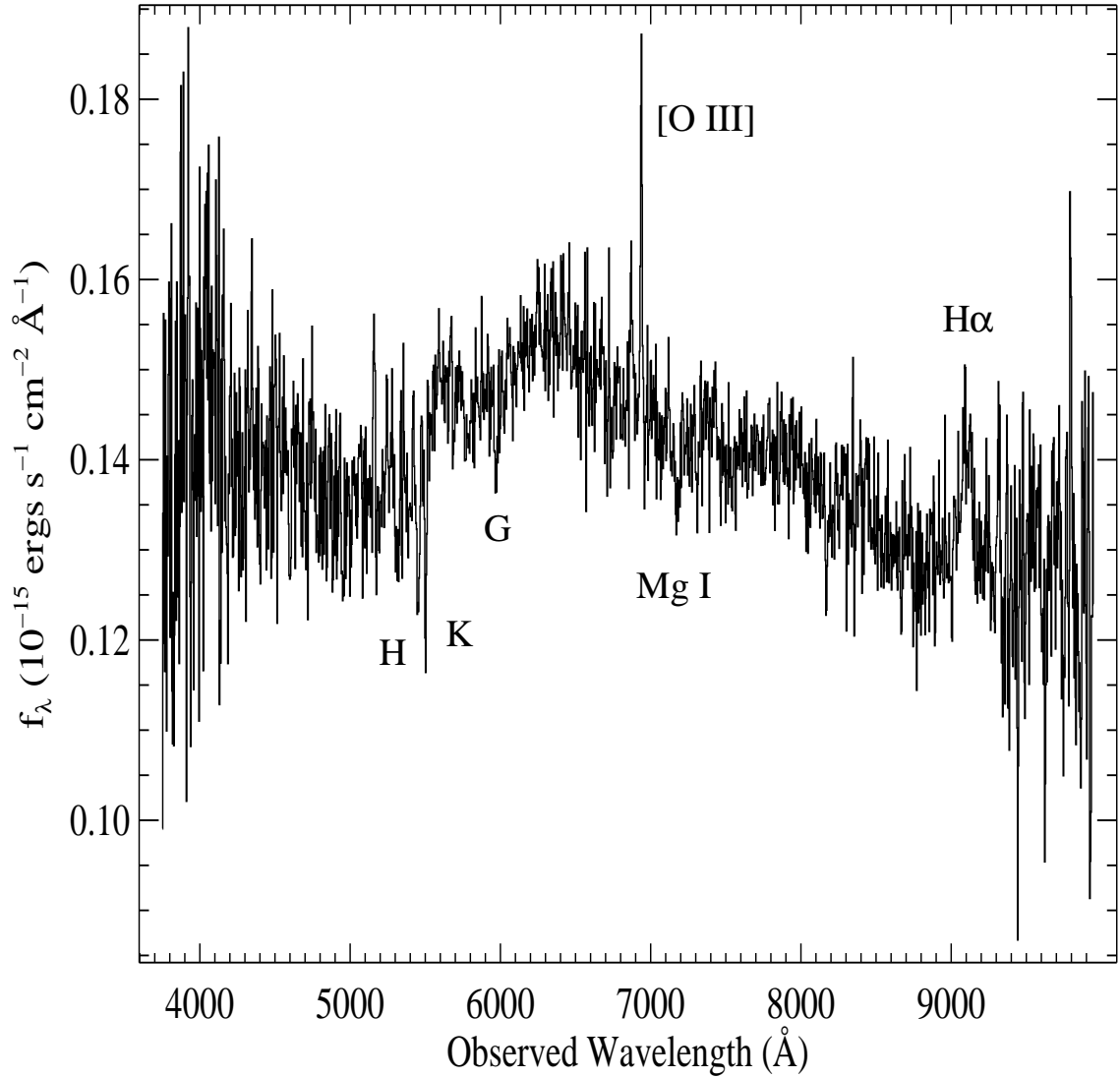}
\caption{A spectrum of SDSS J124602.54+011318.8 obtained on 17.7 January 2002
with the Keck-I 10-m telescope. Note the broad H$\alpha$ emission line and the
emergence of the stellar absorption lines.}
\end{figure*}





\end{document}